# Ten Simple Rules for Creating Biomolecular Graphics

One need only compare the number of three-dimensional molecular illustrations in the first (1990) and third (2004) editions of Voet & Voet's *Biochemistry* in order to appreciate this field's profound communicative value in modern biological sciences – ranging from medicine, physiology, and cell biology, to pharmaceutical chemistry and drug design, to structural and computational biology. The cliché about a picture being worth a thousand words is quite poignant here: The information 'content' of an effectively-constructed piece of molecular graphics can be immense. Because biological function arises from structure, it is difficult to overemphasize the utility of visualization and graphics in molding our current understanding of the molecular nature of biological systems. Nevertheless, creating *effective* molecular graphics is not easy – neither conceptually, nor in terms of effort required. The present collection of *Rules* is meant as a guide for those embarking upon their first molecular illustrations; it most closely parallels the previous collections devoted to publishing papers [1], making oral presentations [2], and creating good posters [3].

### Rule 1: Study the masters; be multidisciplinary.
Molecular graphics has benefited from a rich history of masterful illustrators, pioneers who thrived in the field's confluence of science & art. Examples include the late Irving Geis, who entered from a background in architecture and graphical design, and who's pioneering artistry set him apart as an indisputable master in the first generation (1970s → 1990s) of biomolecular graphics and structure visualization [4]; indeed, Geis' legacy is such that the collection of his work has been purchased by HHMI, and much of it now adorns the walls of various biomedical institutions. The subsequent generation includes such luminaries as David Goodsell, who draws from a background in computational and structural molecular biology. In general, many research areas of great potential relevance to molecular graphics may be only tangentially related to traditional biosciences; these include the theories of statistical graphical design and data / information visualization, as considerably advanced by the likes of Edward Tufte [5] and the late John Tukey. In addition to studying the masters of biomolecular graphics, be adventurous and sample these other areas too.

### Rule 2: Emulate the masters; be opportunistic.
In a nutshell, Rule 1 can be simply rephrased: Read Tufte's books, study Geis' molecular artistry, analyze Goodsell's style of representation, and so on. They provide invaluable lessons and tangible samples of what to strive towards. A useful starting point in creating your own molecular graphics is to begin by emulating the principles illustrated by the works of these masters; selectively choose and incorporate elements of their design patterns in constructing your own graphics. Also, note that a central principle of efficiency is "don't reinvent the wheel". An extension of this idea is that of scavenging: If you see something you like from an entirely unrelated field (*e.g.*, a data representation style from statistics), don't hesitate to simply adapt that for good use in creating your own molecular visualization.

### Rule 3: Transcend the masters; think outside the box.
Nothing is perfect, and there is always room for improvement. This dictum also applies to the state-of-the-art in molecular graphics. The uppermost echelon of molecular artists consists of those who push the boundaries by developing their own characteristic 'style' (Geis, Goodsell). Witness, for instance, the non-photorealistic 'outline' rendering style that is a hallmark of Goodsell's images; indeed, that has proven to be such an effective visualization strategy that it has been introduced into some popular molecular graphics and visualization software packages. In the same spirit of pushing the *status quo*, imagine where protein visualization would be without Jane Richardson's invention of ribbon diagrams [6]. Do not be intimidated by this rule: there still lies *a lot* of room for creativity and ingenuity between the extremes of routine user ↔ graphical guru.

### Rule 4: Be clear; don't obfuscate.
Clarity is a virtue, both in principle and in practice. Just as lucid scientific writing is characterized by streams of sentences woven together such that there is a logical flow to the prose (with readily discernable main points), extracting information from a graphical illustration should not require undue effort or detective work by the reader, and it should not turn into an exercise in deductive logic. Achieving this requires clarity at the level of raw image construction (*e.g.*, choosing what portion of a protein•ligand complex to show, how best to represent atomic interactions in the binding site region), as well as in the accompanying legends and captions that describe the imagery (see *Rule 7*). A corollary of this rule is to be logical and consistent in creating figures. For instance, a manuscript featuring several intricate 3D renditions will be more user-friendly if a 'canonical' perspective is defined early on, and relative orientations are subsequently defined with respect to that. Similarly, introduce clear symbolic and diagrammatic conventions early in the text, and adhere to them throughout.

### Rule 5: Practice patience, and plan ahead.
Cleverly-crafted molecular images are an incredible form of information compression – a single well thought out figure can convey more meaning than pages of text. Thus, in writing a manuscript, place just as high a premium on the quality and clarity of the scientific illustrations as the prose. For instance, a possible rule of thumb is that at least as much time should be spent constructing each figure as is spent writing two pages of single-spaced text (assuming ~500 words/page). First-rate molecular graphics are the cornerstone of many high-quality publications, and there is no royal road to constructing these. It requires considerable patience and planning to avoid 'throwing-together' figures at the last minute to meet a deadline.

### Rule 6: Embrace state of the art tools… and don't be afraid to invent new ones
Rather than stick with old, well-established tools simply because they are familiar or convenient (what your labmate showed you how to use a few years ago), learn early on to embrace new tools and methods. The initial effort you expend in learning a feature-rich software package (PyMOL, VMD, *etc.*) will be repaid manyfold once you've scaled the learning curve and are able to generate images that are both scientifically compelling and aesthetically appealing. This *Rule* applies to both software *and* hardware. The software stage encompasses everything from raw image creation and rendering (*e.g.*, ray tracing) to final graphical layout (*e.g.*, using a vector graphics tool like Inkscape or Adobe Illustrator). On the hardware front, advances continue at an astonishing rate. Embrace the latest technologies, such as nVidia's 'Gelato' (for GPU-accelerated graphical rendering), learn about sophisticated methodologies like ambient occlusion lighting, and so on. Finally, if the data complexity ('representation pressure') demands it, try implementing or designing new tools – novel representation styles, diagrammatic conventions, software, *etc*. This "dig into the code" philosophy is a corollary to Rules 3 & 6; to be poised to act on it, note that you will vastly expand your graphical horizons by learning programming or scripting languages compatible with the application programming interface (API) of your favorite graphics packages (if PyMOL then Python, if VMD then Python or Tcl, *etc.*). Be careful with this Rule, though: Overdoing it could lead to an aesthetically pleasing but scientifically hollow (or at least overstated) illustration.

### Rule 7: Captions, captions, captions.
Captions should not be overlooked or given short-shrift. A well-written caption that accompanies a useless graphical panel will likely come across as an afterthought. Conversely, a beautiful, information-rich image lacking a correspondingly high–quality caption is hardly more informative than a random array of pixels. Therefore, production of first-rate biomolecular graphics requires that sufficient effort be dedicated to this often-overlooked part of the figure. Indeed, some authors conquer writer's block by first organizing the data and results into clear, compelling graphical panels, and then filling-out these panels with over-long captions which can then be co-opted into the body of the nascent manuscript.

### Rule 8: Have others critique your illustrations.
Ask your most ruthlessly critical colleagues to peruse your figures (even while they are still works in progress), with an eye towards what can be improved, what can be trimmed or considered superfluous, what may be missing, and so on. Following through on this vital step will enhance the pedagogical value of your illustration, making it useful to those aside from you. Doing so earlier rather than later will avoid potentially wasting time on what is shaping-up to be an obtuse or unclear figure.

### Rule 9: Notice bad examples (and what makes them bad).
In perusing the literature, take special note when you run across particularly bad figures, unclear artwork, poorly-designed illustrations, opaque captions, *etc.*; most importantly, carefully analyze these figures to pinpoint what you consider to be their shortcomings, and strive to avoid such pitfalls in your own work. Note that this parallels *Rule 1* of ref. [1].

### Rule 10: Tailor to the task or audience at hand.
Most of the molecular graphics in a definitive, 20-page tome in some subject area will not be suitable for a far more concise, five-page manuscript meant for a general audience. Similarly, a manuscript's graphics and figures will likely need to be reformulated (or at least somewhat tailored) for effective use in the context of a poster or oral presentation, versus a manuscript. Though this may seem like an unnecessary waste of time, it stems from the basic principle that one should optimally match illustrations and graphical work with the intended audience and purpose; fortunately, the burden of doing so will diminish as you gradually build-up a library of raw images and figure panels which can then be rearranged in generating new slides, posters, *etc.* In the same vein, one should bear in mind certain 'best practices' so as to maximize the audience to which your graphical artwork is accessible (*e.g.*, employing color charts and texture maps to build alternative sets of images that are interpretable by colorblind viewers).

### Acknowledgements
My molecular graphics education owes much to the UCLA structural biology community (particularly D Anderson, D Cascio, D Eisenberg, & MR Sawaya), who always provided incisive advice on, and instructive examples of, the power of molecular visualization and graphical production.  I thank L Columbus for critical reading of the manuscript.

## Postscript

For a further treatment of this topic, see:

*An Introduction to Biomolecular Graphics*
Mura C, McCrimmon CM, Vertrees J & Sawaya MR.
*PLoS Computational Biology* (2010), *6(8)*: e1000918.
http://dx.doi.org/10.1371/journal.pcbi.1000918 ♦ PMID 20865174